\documentstyle[11pt,epsfig]{article}
\begin{document}
\title{Semi-analytic approximations\\ for production
of atmospheric muons and neutrinos\thanks{This research is supported
in part by the US Department of Energy grant DE-FG02 91ER 40626.}}
\author{Thomas K. Gaisser \\
Bartol Research Institute, University of Delaware\\
Newark, DE 19716}

%%%%%%%%%%%%%%%%%%%%%%%%%%%%%%%%%%%%%%%%%%%%%%%%%%%%%%%%%%%%%%%%%%%%%%%%

\maketitle
\begin{abstract}
Simple approximations for fluxes of atmospheric muons and
muon neutrinos
are developed which display explicitly how the fluxes depend on primary
cosmic ray energy and on features of pion production.  For energies
of approximately 10 GeV and above the results are sufficiently accurate
to calculate response functions and to use for estimates of systematic
uncertainties.
\end{abstract}

%%%%%%%%%%%%%%%%%%%%%%%%%%%%%%%%%%%%%%%%%%%%%%%%%%%%%%%%%%%%%%%%%%%%%%%%
\section{Introduction}
There are several reasons for which it is useful to have
simple approximations for production of muons and neutrinos
by interactions of cosmic-ray nucleons in the atmosphere.
First, the dependence on pion and kaon production, as well as
the dependence on energy of the primary cosmic rays can
be displayed explicitly.  This is useful to make estimates
of the systematic errors in the expected signals in muon and
neutrino detectors that arise from  uncertainties in the
primary cosmic-ray spectrum and in meson production, which
in general depend on energy.

The dependence of the signal in a particular detector
on the primary energy spectrum is its response function.
The classical use of response functions for muon telescopes~\cite{TGresponse}
is to interpret measurements of solar modulation of muons
as a function of muon energy and geomagnetic latitude~\cite{mumod}.
In addition, response functions are useful for planning and optimizing
Monte Carlo
simulations of detectors because they display explicitly the relative
portion of the signal that comes from each part of the primary spectrum.

For power-law spectra the analytic approximations for fluxes of muons
and neutrinos are well-known~\cite{early,Gaisser,Lipari}.
In this paper we develop approximations which also display
the dependence on primary energy.  Such approximations
allow semi-analytic evaluation of response functions.  They can also
be used to estimate the muon-induced signal produced by a fluence of protons
with an arbitrary energy spectrum and time profile,
such as might be associated with a strong solar event.

The approximate results here are valid for muons and muon-neutrinos
at relatively high energy, of order ten GeV and above.  This includes
the interesting cases of underground measurements of atmospheric muons
and of upward energetic muons induced by atmospheric neutrinos.
The same results can, of course, be obtained with greater precision
and for low energies, including electron neutrinos,
by Monte Carlo simulations or by more complicated numerical
calculations.  The analytic approximations are nevertheless useful
for the insight they provide and for the diagnostic purposes mentioned above.

\section{Derivation of analytic approximations}

The differential production spectra of muons and neutrinos
of energy $E$ from decay of pions in atmospheric cosmic ray cascades are
given by
\begin{eqnarray}
\label{production}
{dN\over dE\,dE_\pi dE_N dX}&=&{S(E,E_\pi)\over (1-r_\pi)E_\pi}
{1\over d_\pi}\,\int_0^X\,{dX^\prime\over \lambda_N}\,
P_\pi(E_\pi,X,X^\prime)\\ \nonumber
& &\times {1\over E_\pi}F_{N\pi}(E_\pi,E_N)\,
e^{-X^\prime/\Lambda_N}\,\phi(E_N),
\end{eqnarray}
where $P_\pi(E_\pi,X,X^\prime)$ is the probability that a charged pion
produced at slant depth $X^\prime$~(g/cm$^2$) survives to depth
$X\,(>\,X^\prime)$.
For muons the kinematic factor, $S(E,E_\pi)$, has the form
\begin{equation}
S_\mu(E_{\mu,0},E_\pi)\,=\,\Theta(E_\pi-E_{\mu,0})\,
\Theta({E_{\mu,0}\over r_\pi}-E_\pi),
\label{kinmu}
\end{equation}
while for muon neutrinos it is
\begin{equation}
S_{\nu_\mu}(E_\nu,E_\pi)\,=\,\Theta(E_\pi-{E_\nu\over (1-r_\pi)}),
\label{kinnu}
\end{equation}
where $r_\pi\,=\,{m_\mu^2\over m_\pi^2}$.  Here
$\Theta$ is the Heaviside step function, and
the subscript~$_0$ in Eq.~\ref{kinmu}
indicates muon energy at production (see Appendix).

The contribution from kaon-decay has the same form as Eq.~\ref{production}
with appropriate changes in kinematic factors and an overall
factor for the branching ratio of $B_{K\rightarrow \mu\nu}\,=\,0.635$.
The various parameters
relevant for $N\rightarrow[K,\pi]\rightarrow \mu\nu_\mu$
are summarized in Table 1.  The differences between the decay parameters
for kaons and pions and between the mass ratios, $r_\pi$ and $r_K$
are responsible for the striking difference between the importance
of kaons for neutrinos as compared to muons. In particular,
kaons become the dominant source of atmospheric
neutrinos for $E_\nu$ larger than
$\sim100$~GeV even though more pions than kaons are produced.

\begin{table}
\caption{Parameters for atmospheric $\mu^+ +\mu^-$ and $\nu_\mu +\bar{\nu}_\mu$}
\centerline{
\begin{tabular}{|lllll|}\hline
Mass-square ratios:&$r_\pi$&$r_K$&& \\
&0.573 & .046 & &\\ \hline
Characteristic $E_{decay}$:&$\epsilon_\pi$&$\epsilon_K$&$\epsilon_\mu$& \\
&115 GeV & 850 GeV & 1 GeV & \\ \hline
Attenuation lengths:&$\Lambda_\pi$&$\Lambda_K$&$\Lambda_N$&$\lambda_N$ \\
&160 g/cm$^2$ & 180 g/cm$^2$ & 120 g/cm$^2$ & 86 g/cm$^2$ \\ \hline
\end{tabular}
}
\label{tab1}
\end{table}

The integral in Eq.~\ref{production} is over atmospheric depth $X^\prime$
measured in g/cm$^2$ along the trajectory of the incident particle
at zenith angle $\theta$.  The nucleon interaction length is
$\lambda_N$.  The factor
$\exp\{-X^\prime/\Lambda_N\}\,\phi(E_N)$ is the flux of nucleons
of energy $E_N$ at slant depth $X^\prime$, where $\Lambda_N$ is
the nucleon attenuation length in the atmosphere and $\phi(E_N)$ is
the primary spectrum of nucleons per GeV/nucleon.  The superposition
approximation~\cite{JEngel} is used for nucleons bound in nuclei.

The normalized inclusive cross section
for $N\,+\,{\rm air}\rightarrow \pi^\pm\,+\,X$ is
\begin{equation}
F_{N\pi}(E_\pi,E_N)
\,=\,{E_\pi\over\sigma_N}\,{d\sigma(E_\pi,E_N)\over dE_\pi}\approx
c_+(1-x)^{p_+}\,+\,c_-(1-x)^{p_-},
\label{Zpion}
\end{equation}
where $x=x_\pi=E_\pi/E_N$.  The parameters of this scaling approximation
are fixed by comparison to accelerator data in the sub-TeV
range~\cite{Gaisser}, and the adequacy of the simple approximation
is verified later by comparison of the calculated neutrino flux to
the result of a complete Monte Carlo calculation.  A more complicated
or precise expression could be used for the inclusive cross section
if desired.  The expressions for kaons have the same form, and the
numerical values used in this paper are given in Table 2.

\begin{table}
\caption{Parameters for kaon and pion production}
\centerline{
\begin{tabular}{l|cccccc}
&$c_+$&$p_+$&$Z_+$&$c_-$&$p_-$&$Z_-$ \\ \hline
charged pions & 0.92&4.1&0.047&0.81&4.8&0.034 \\
charged kaons & 0.037 & 0.87& 0.0089 & 0.045 & 3.5 & 0.0028 \\ \hline
\end{tabular}
}
\label{tab2}
\end{table}

The survival probability in Eq.~\ref{production} is a product of
survival against decay and survival against interaction.
\begin{equation}
P_\pi(E_\pi,X,X^\prime)\;=\;
\left({X^\prime\over X}\right)^{\epsilon_\pi/E_\pi\cos\theta}\,
\exp\{-(X-X^\prime)/\Lambda_\pi\}.
\label{surv_pi}
\end{equation}
Use of the pion attenuation length, $\Lambda_\pi$, rather than the
pion interaction length, $\lambda_\pi < \Lambda_\pi$, in
the interaction factor of Eq.~\ref{surv_pi}
accounts approximately for the regeneration of charged pions when
pions interact in the atmosphere.
The first factor in Eq.~\ref{surv_pi}
is the probability for survival against decay, and the explicit
form used here is valid in the approximation of an exponential
atmosphere with scale height $h_0$ and decay constant
$\epsilon_\pi=m_\pi c^2h_0/c\tau_\pi$.

Finally, $1/d_\pi\,=\,\epsilon_\pi/(E_\pi X\cos\theta)$ is the
differential pion decay probability (per g/cm$^2$) in an
exponential atmosphere.  The simple angular dependence displayed
here is valid when the curvature of the Earth can be neglected
($\theta \le 60^\circ$).  For large angle the analysis is more
complicated, but one approach is to use the same form with
$\cos\theta$ replaced by an effective value that depends on
$\theta$~\cite{pathlength}.

The next step is to carry out the integral over the production depth,
$X^\prime$, in Eq.~\ref{production}.  This is done by expanding the
exponential of $X^\prime$ and expressing the result as a series
of the form
\begin{equation}
{1\over d_\pi}\,\int_0^X\,P_\pi(E_\pi,X,X^\prime)
\,e^{-X^\prime/\Lambda_N}\,{dX^\prime\over
\lambda_N}\,=\,{e^{-X/\Lambda_\pi}\over\lambda_N}\,
\sum_{n=1}^\infty \left({X\over\Lambda_\pi}\,(-a_\pi)\right)^{n-1}\,
{1\over n!(1+nE_\pi\cos\theta/\epsilon_\pi)},
\label{partial}
\end{equation}
where $a_\pi\,=\,\Lambda_\pi/\Lambda_N\,-1\,\approx\,0.33$.
Substituting this expression into Eq.~\ref{production} we can now integrate
over all depths $X$ of pion decay.  The upper limit of this integral
can safely be extended to infinity since most production of secondaries
occurs well above the ground.
The result for $\nu_\mu\,+\,\bar{\nu}_\mu$ is
\begin{equation}
{dN\over dE_\nu\,dE_\pi dE_N}\,=\,{S_\nu(E_\nu,E_\pi)\over (1-r_\pi)E_\pi}\,
{1\over E_\pi}F_{N\pi}(E_\pi,E_N)\,G_\pi(E_\pi,\theta)\,\phi(E_N).
\label{production2}
\end{equation}

The physics of cascade propagation is included in Eq.~\ref{production2} by
the factor
\begin{equation}
G_{\pi}(E_\pi,\theta)\;=\;{\Lambda_\pi\over \lambda_N}\,\sum_{n=1}^\infty\,
\left\{ {(-a_\pi)^{n-1}\over 1+n\,E_\pi\,\cos\theta/\epsilon_\pi}\right\}.
\label{Gpions}
\end{equation}
The contribution from kaons has the same form with
$a_K\,=\,\Lambda_K/\Lambda_N\,-\,1\,\approx\,0.5$.

The dependence on the inclusive cross sections
for pion (and kaon) production by nucleons can be displayed
immediately in the approximation of a power-law primary spectrum
by changing variables from $\{E_\nu,E_\pi,E_N\}$
to $\{E_\nu,x_\pi,z=x_\pi E_N\}$ and integrating over $z$ to obtain
\begin{equation}
{dN\over dE_\nu dx_\pi}\;=\;\left(x_\pi\right)^{\gamma-1}{F_{N\pi}(x_\pi)\over
1-r_\pi}\,\int_{E_\nu/(1-r_\pi)}^\infty\,{dz\over z}\,G_\pi(z,\theta)\,\phi(z),
\label{Zform}
\end{equation}
where $\gamma$ is the integral spectral index.
The dependence on pion production by nucleons appears here as an explicit
proportionality factor, the integral of which is the Z-factor,
\begin{equation}
Z_{N\rightarrow\pi}\;=\;\int_0^1\,dx\,(x)^{\gamma - 1}\,
F_{N\rightarrow\pi}(x),\;{\rm etc.}.
\label{Zfactor}
\end{equation}
The dependence on nucleon production by nucleons and on pion [kaon]
production by pions [kaons] does not appear explicitly in Eq.~\ref{Zform}.
Dependence on these processes is contained
in the attenuation lengths, $\Lambda_N$ and $\Lambda_\pi$ [$\Lambda_K$].
which appear respectively in Eq.~\ref{production} and in the expression
for $G_\pi$ [$G_K$] as defined in Eq.~\ref{Gpions}.  The attenuation
lengths are related to the respective interaction lengths and
inclusive cross sections by
$\Lambda_i\,=\,\lambda_i/(1\,-\,Z_{i\rightarrow i})$, where $i$ stands
for nucleon (N = p or n), charged pion or charged kaon.
Note from Eq.~\ref{Gpions} that $\Lambda_\pi$ cancels out of the
expression when $E_\pi/\epsilon_\pi\ll 1$, reflecting the fact that
low energy pions [kaons] usually decay before they interact.
The small cross processes $K^\pm\leftrightarrow\pi^\pm$,
as well as neutral kaons,
are neglected as sources of $\nu$ and $\mu$ in these approximations.

\begin{figure}[htb]
\flushleft{\epsfig{figure=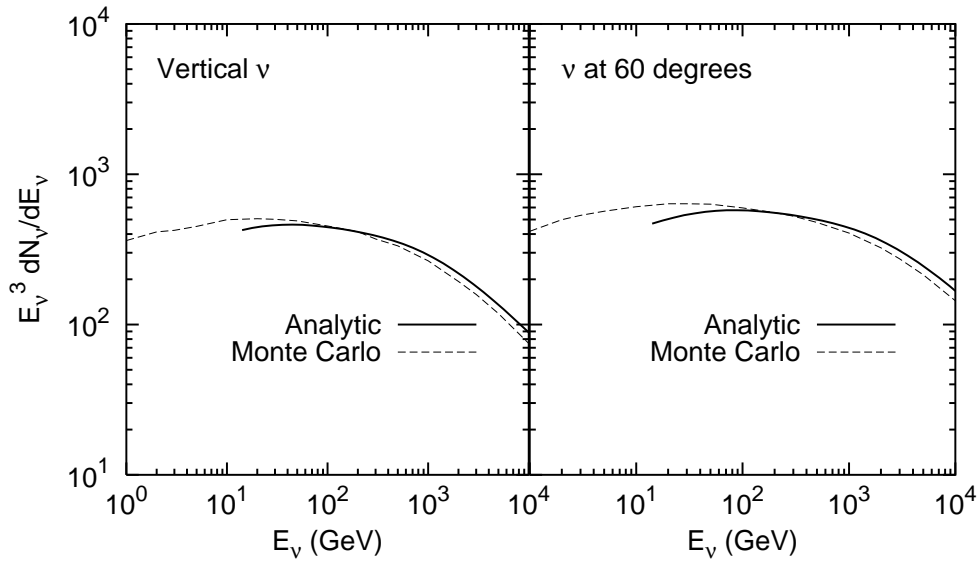,width=14.5cm}}
\caption{Comparison of the analytic approximation (solid)
for the flux of $\nu_\mu+\bar{\nu}_\mu$ with the result of the full
Monte Carlo (dashed)~\protect\cite{AGLS}.
}
\label{Nu}
\end{figure}

\begin{figure}[htb]
\flushleft{\epsfig{figure=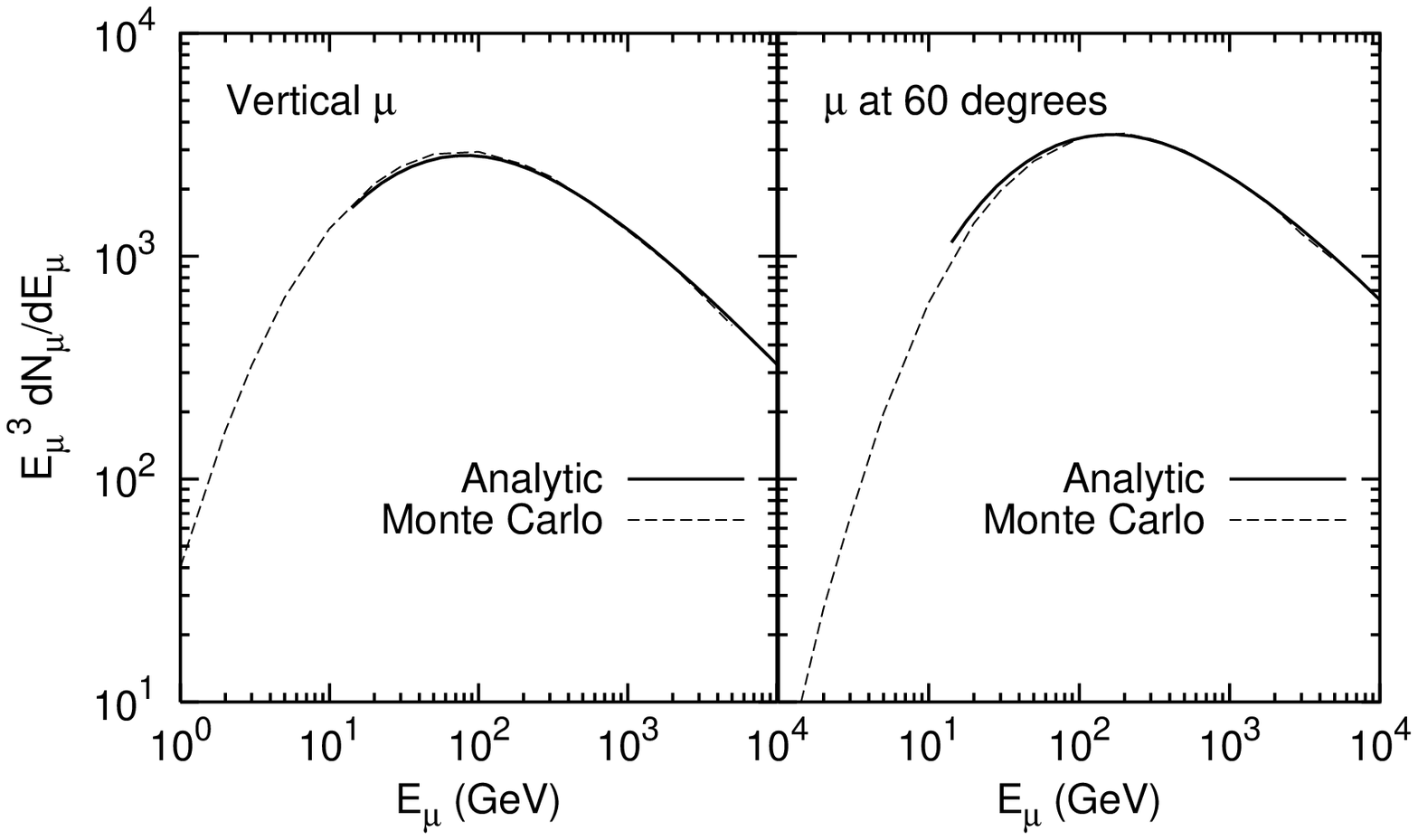,width=14.5cm}}
\caption{Comparison of the analytic approximation (solid)
for the flux of $\mu^+ +\mu^-$ with the result of the full
Monte Carlo (dashed)~\protect\cite{AGLS}.
}
\label{Mu}
\end{figure}

\section{Response Functions and Fluxes}
To obtain the response function,
the integral over pion energy is carried out in Eq.~\ref{production2}
making use of the scaling approximation of Eq.~\ref{Zpion}.
The flux of $\nu_\mu+\bar{\nu}_\mu$ with zenith angle $\theta$
from decay of $\pi^+ +\pi^-$ is given by
\begin{equation}
E_N\left.{dN_\nu\over dE_\nu dE_N}\right|_\pi\;=\;{\phi(E_N)\over 1 - r_\pi}
\int_{x_\nu/(1-r_\pi)}^1\,dx\,G_{\pi}(E_Nx,\theta)\,{F_\pi(x)\over x^2},
\label{pions}
\end{equation}
where $x_\nu\,=\,E_\nu/E_N$.  The contribution from kaons has the same form.

In the absence of muon decay and energy loss, the expression for the
response function for muons would be identical to that for neutrinos
except that the limits of the integration in Eq.~\ref{pions} are changed
to those appropriate for muons (Eq.~\ref{kinmu} instead of \ref{kinnu}).
They are
$$ x_{min}\,=\, {E_{\mu,0}\over E_N}\;\;{\rm and}\;
x_{max} \,=\,\min\left[\,1,\,{x_{min}\over r_\pi}\,\right].
$$
In the Appendix an approximate treatment is given for the effects of muon
decay and energy loss that can be used for $E_\mu\sim 10$~GeV and higher.
In this approximation,
$$
E_{\mu,0}\,\approx\,\left(E_\mu + {2\,{\rm GeV}\over\cos\theta}\right),
$$
and
\begin{equation}
E_N\left.{dN_\mu\over dE_\mu dE_N}\right|_\pi\;=\;{\phi(E_N)\over 1 - r_\pi}\,
A(E_\mu)\,\int_{x_{min}}^{x_{max}}\,dx\,G_{\pi,\mu}(E_Nx,\theta)\,
{F_\pi(x)\over x^2},
\label{muons}
\end{equation}
with
\begin{equation}
G_{\pi,\mu}(E_Nx,\theta)\;=\;{\Lambda_\pi\over \lambda_N}\,\sum_{n=1}^\infty\,
C_n(E_\mu)\,\left\{ {(-a_\pi)^{n-1}\over
1+nE_N\,x\,|\cos\theta|/\epsilon_\pi}\right\}.
\label{Gmuons}
\end{equation}
The functions $A(E_\mu)$ and $C_n(E_\mu)$ are defined in the Appendix.

\begin{figure}[htb]
\flushleft{\epsfig{figure=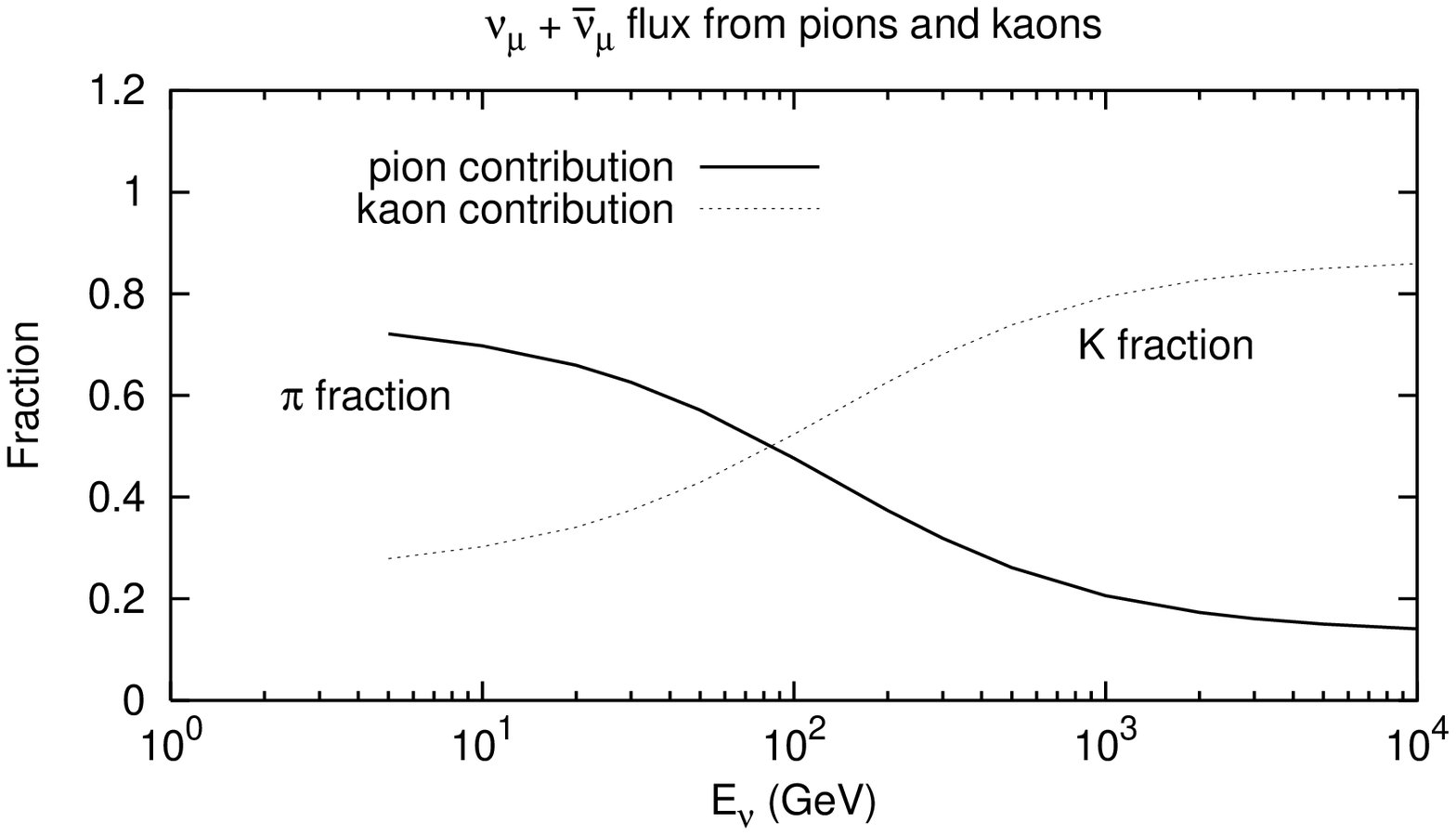,width=14.5cm}}
\caption{Plot showing fractional contribution of pions and
kaons to the flux of $\nu_\mu+\bar{\nu}_\mu$.
}
\label{Fig1}
\end{figure}

The differential fluxes are then obtained in both cases
by summing the contributions from pions and kaons and integrating
over $E_N$.
The analytic approximations for the fluxes of $\nu_\mu+\bar{\nu}_\mu$
and $\mu^+ +\mu^-$ are compared with the results of the full
Monte Carlo calculation~\cite{AGLS} at two angles in Figs~\ref{Nu}
and~\ref{Mu}.  The values of parameters that have been used in
the analytic approximations are listed in Tables 1 and 2.  They correspond
most closely to the hadroproduction model of Ref.~\cite{AGLS} for
interaction energies around 100 GeV.  The approximation falls below the neutrino
flux at low energy primarily because the contribution from muon
decay is not included in the approximation.  At high energy the
approximation is 10-15\% higher than the Monte Carlo, because
the simple, energy-independent 
forms for kaon production used here are not identical to the
actual representation in the Monte Carlo.   The analytic approximations
could be tuned to give a better representation of this~\cite{AGLS} or
any other interaction model, but doing so is not the object of this
paper.  The approximate results are sufficiently accurate 
that the analytic calculation may be used for diagnostic purposes
above 10~GeV where the
contributions from muon decay are unimportant. 
These include propagation of errors,
calculation of response functions and evaluation of the
phase space regions that are most important for the
determining the fluxes of neutrinos and muons.

In Fig.~\ref{Fig1} I show the fractions of
$\nu_\mu+\bar{\nu}_\mu$ from $\pi^\pm$ and from $K^\pm$ to demonstrate
the importance of kaon production.
Note that over the entire energy region, approximately
70-75\% of $\phi_\nu$ comes from free protons (most of the rest
is from nucleons in primary helium)~\cite{GS}.  Thus, any uncertainty
from the approximate, superposition model treatment~\cite{JEngel}
of bound nucleons will be small.

\begin{figure}[htb]
\flushleft{\epsfig{figure=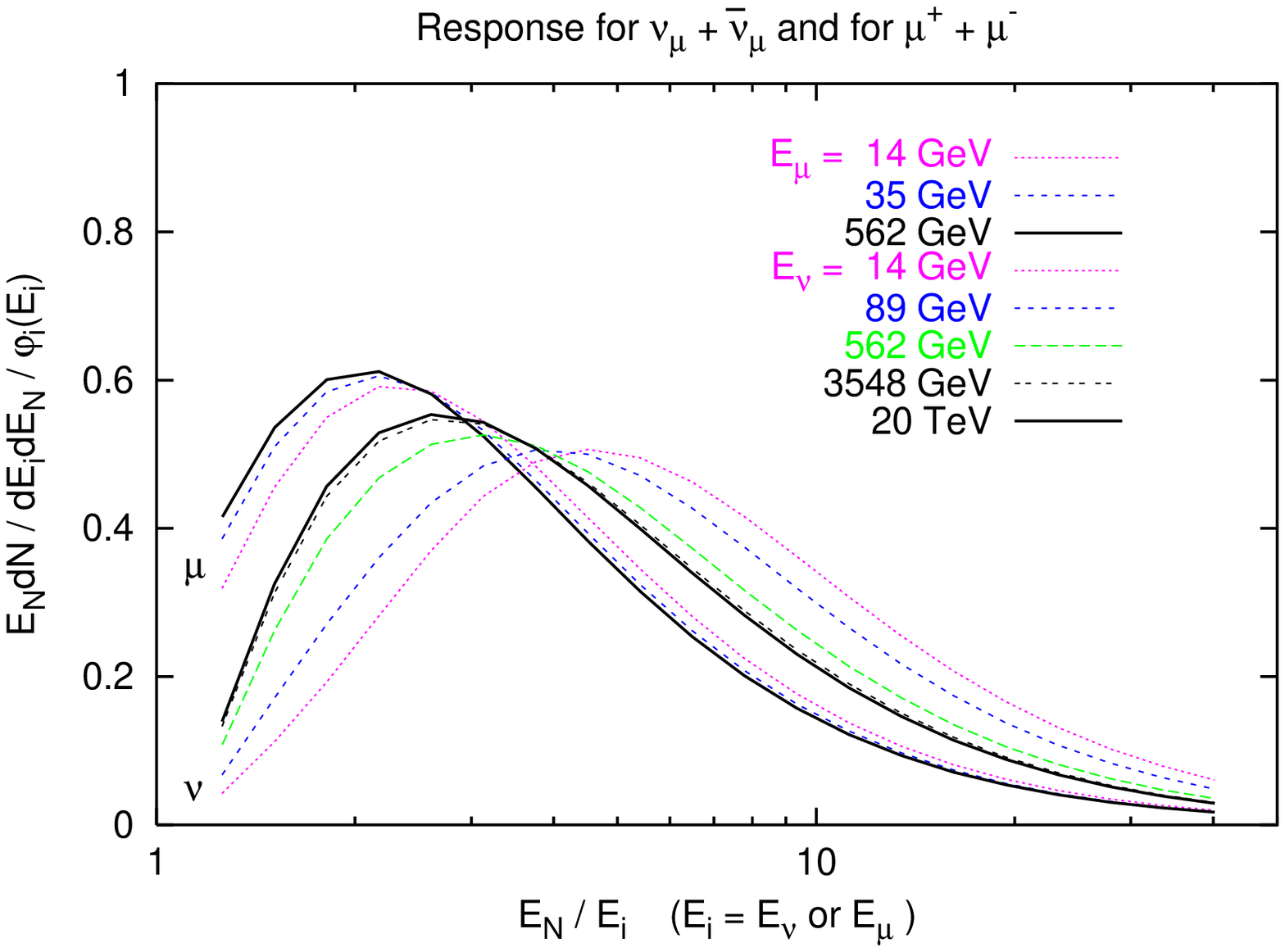,width=14.5cm}}
\caption{The distributions of primary energy per nucleon ($E_N$) that
give rise to vertical $\nu_\mu+\bar{\nu}_\mu$ of energy $E_\nu$
(right-hand set of curves) and to $\mu^+\,+\,\mu^-$.
Each curve is normalized to unity.  For each set the lowest
energy is the rightmost curve.  The response is approximately
energy-independent for $E_\nu>3$~TeV and for $E_\mu> 30$~GeV
when expressed in terms of $E/E_N$.}
\label{Fig2}
\end{figure}

Having established that the method is adequate, the next step is
to calculate the response functions for a range of energies.
This is illustrated in Fig.~\ref{Fig2} for muons and
neutrinos.  Note that the neutrino response is shifted to
higher primary energy than the response for muons.  In addition,
it is slower to reach an asymptotic form.  These features are
consequences of the kinematics of $\pi\rightarrow\mu\nu_\mu$ as
compared to $K\rightarrow\mu\nu_\mu$ coupled with the greater
relative importance of kaons for neutrinos at high energy.
Pions must have more than twice the energy of daughter $\nu_\mu$,
while $E_\pi\ge E_{\mu,0}$ (compare Eqs.\ref{kinmu},\ref{kinnu}).
This shifts the $\nu_\mu$ response to higher energy when
the pion chain is dominant.  At higher energy, where kaons dominate
the production of $\nu_\mu$ the response shifts to lower energy.

\begin{figure}[htb]
\flushleft{\epsfig{figure=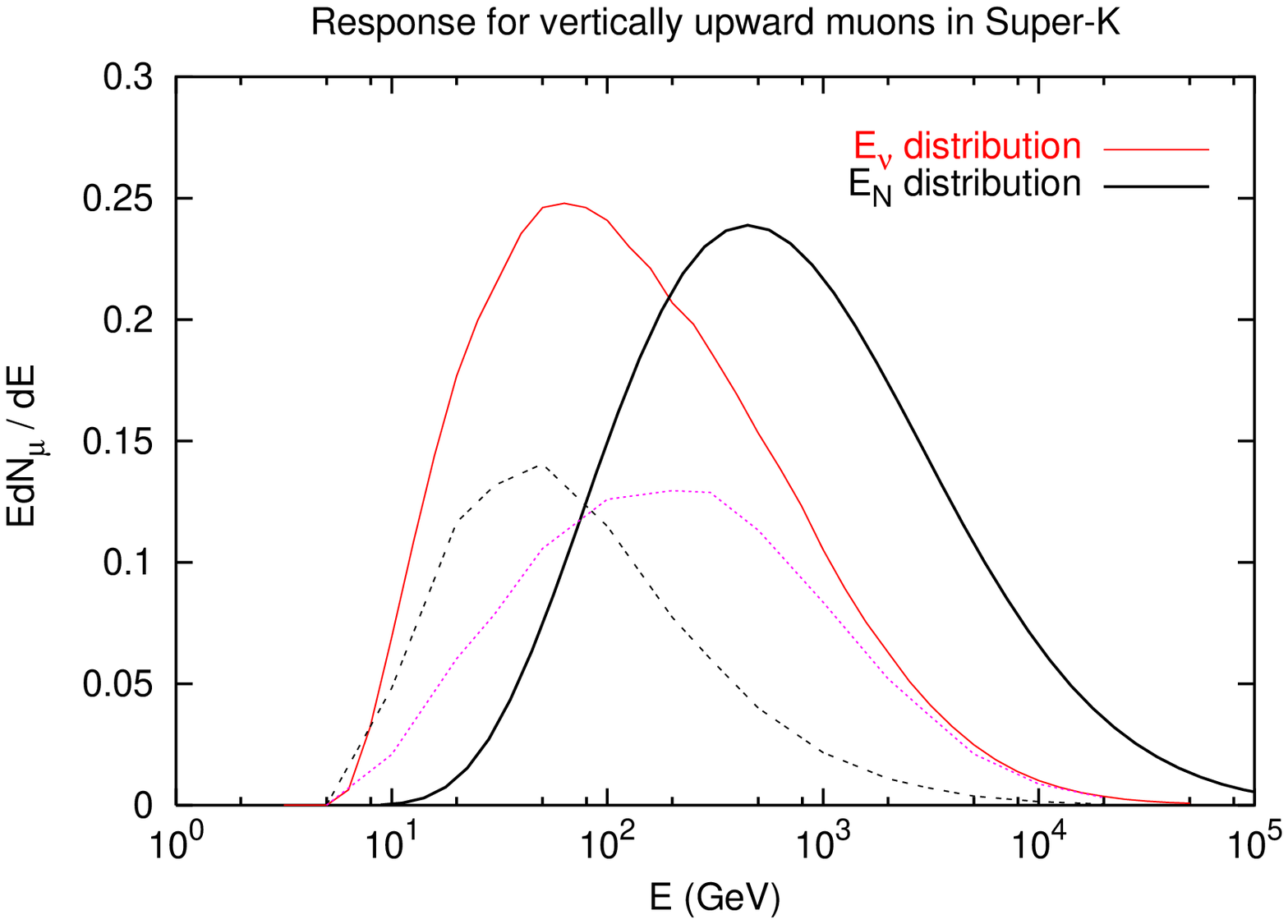,width=14cm}}
\caption{Distribution of neutrino energy (light solid) and primary
nucleon energy (heavy solid) that produce the through-going, vertically
upward signal at Super-Kamiokande.  Each curve is normalized to unity.
The contributions of pions (double dash) and of kaons (dotted) are shown
as a function of neutrino energy.
}
\label{resp-sk}
\end{figure}

\section{Estimates of Uncertainties}
To illustrate how the formalism presented here can be used to
estimate uncertainties in the calculated result, we consider the
example of vertically upward, neutrino-induced, through-going muons
at Super-Kamiokande~\cite{SuperK}.  The distribution of neutrino energies that
produce this signal (in the absence of oscillations)
is shown as the light solid line~\cite{Ralph} in
Fig.~\ref{resp-sk}.  The heavy line shows the corresponding distribution of
parent nucleon energies.  This distribution should be used to
weight the uncertainties in the input in a convolution over primary
energy.  A similar analysis can be made for neutrino-induced muons
that stop inside the detector.

There are two main sources of uncertainties in calculations
of fluxes of muons and neutrinos with energies in the multi-GeV range
and above.  We deal first with those arising from uncertainties in
the primary spectrum and then with those from uncertainties in treatment
of pion and kaon production.

\subsection{Primary spectrum}
\begin{figure}[htb]
\flushleft{\epsfig{figure=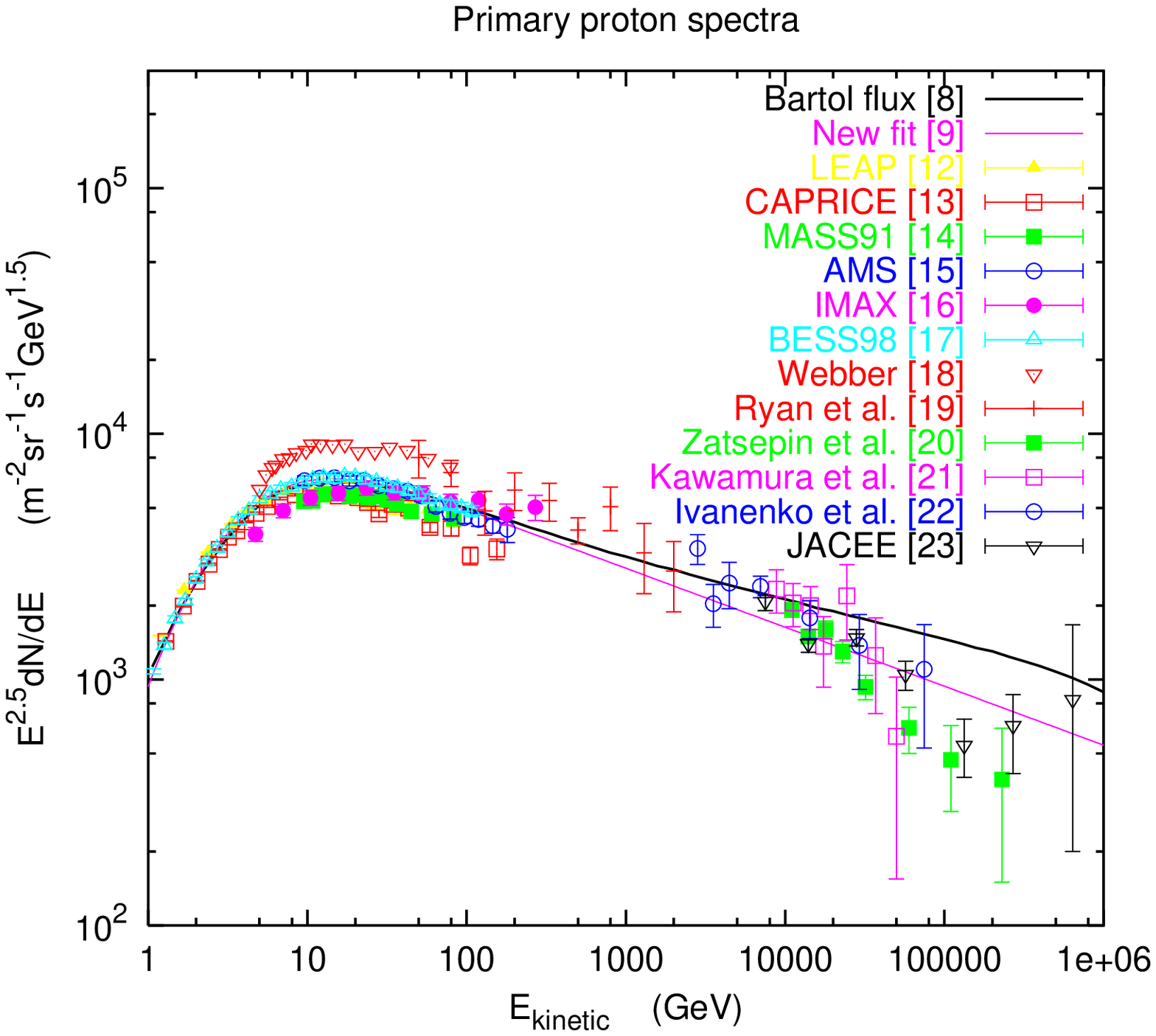,width=14cm}}
\caption{Summary of measurements of the spectrum
of cosmic-ray protons.  Data are
from~\protect\cite{LEAP,CAP,M91,AMS,IMAX,B98,W,Ryan,Zat,Kaw,Ivan,JAC}.
}
\label{protons}
\end{figure}

Since approximately 75\% of high energy atmospheric neutrinos
are produced by incident protons, I concentrate on that component of the
cosmic-ray spectrum for the analysis of uncertainties.  A summary of data is
shown in Fig.~\ref{protons}.  The solid line is the proton
spectrum used in Ref.~\cite{AGLS}, which I take as the reference
spectrum.  The resulting uncertainty in the flux
of atmospheric neutrinos that is relevant for the Super-K example
will be a convolution of the uncertainties
in the data of Fig.~\ref{protons} with the response function shown
as the heavy solid line in Fig.~\ref{resp-sk}.

It is natural to divide the discussion
into three energy regions:

\subsubsection{Up to $\sim100$ GeV}
Primaries with energies below 100 GeV
are the most precisely measured.
They make the major contribution to contained neutrino
events, but only a small contribution to neutrino-induced
upward muons.  While they do not cover the full range of
interest they can be used to renormalize the single
measurement~\cite{Ryan} that spans the important energy
range up to $\sim1$~TeV.  Several measurements have been made
in the past decade with balloon-borne magnetic spectrometers capable
of making relatively precise momentum measurements for well-identified
protons and nuclei.  Since the LEAP experiment~\cite{LEAP},
five other experiments~\cite{CAP,M91,AMS,IMAX,B98} confirm a significantly
lower normalization than the earlier standard reference~\cite{W}.
Particularly noteworthy are the BESS98~\cite{B98} and the AMS~\cite{AMS}
measurements, which agree with each other to within 5\% over two
decades in energy.  These data have been used as the basis of a new
fit~\cite{GS} as shown in Fig.~\ref{protons}.
The systematic uncertainty in the primary spectrum
below 100 GeV is estimated in Ref.~\cite{GS} as $<\pm5$\%.

\subsubsection{Up to $\sim1$ TeV}
The energy region from 0.1 to 2 TeV, explored by the large balloon-borne
tracking calorimeter of Ryan {\it et al.}~\cite{Ryan},
is of great importance for $\nu_\mu$-induced upward muons.
We renormalize the Ryan {\it et al.} data downward by 25 \% to agree with
the magnetic spectrometer data below 100~GeV~\cite{GS}.  Note that
the normalization can reflect a smaller shift in
energy calibration because of the steep spectrum~\cite{GS}.
This renormalization brings the measurement of Ryan {\it et al.} into
agreement with the reference spectrum.  The
remaining overall uncertainty in this energy region is
$\sim\pm 10$\%.

\subsubsection{The multi-TeV region}
Most of the data in this energy range comes from balloon flights of emulsion
chambers,~\cite{Zat,Kaw,JAC} which measure only
the electromagnetic component of the energy
deposited by the interacting proton in the chamber.
This is subject to large fluctuations superimposed
on the uncertainty in the underlying inelasticity distribution
of protons when they interact in the emulsion chamber.
There is also one measurement in this energy region made
with a tracking calorimeter carried on flights of the
SOKOL satellite~\cite{Ivan}.

The uncertainties in these measurements are
at the level of $\pm25$\%.  In addition, there is an indication
that the flux falls below the reference spectrum above $\sim20$~TeV
by up to a factor of 2.  However, the contribution from protons with
$E_N>20$~TeV to the energy range responsible for $\nu_\mu$-induced
upward muons is small.

\begin{figure}[htb]
\flushleft{\epsfig{figure=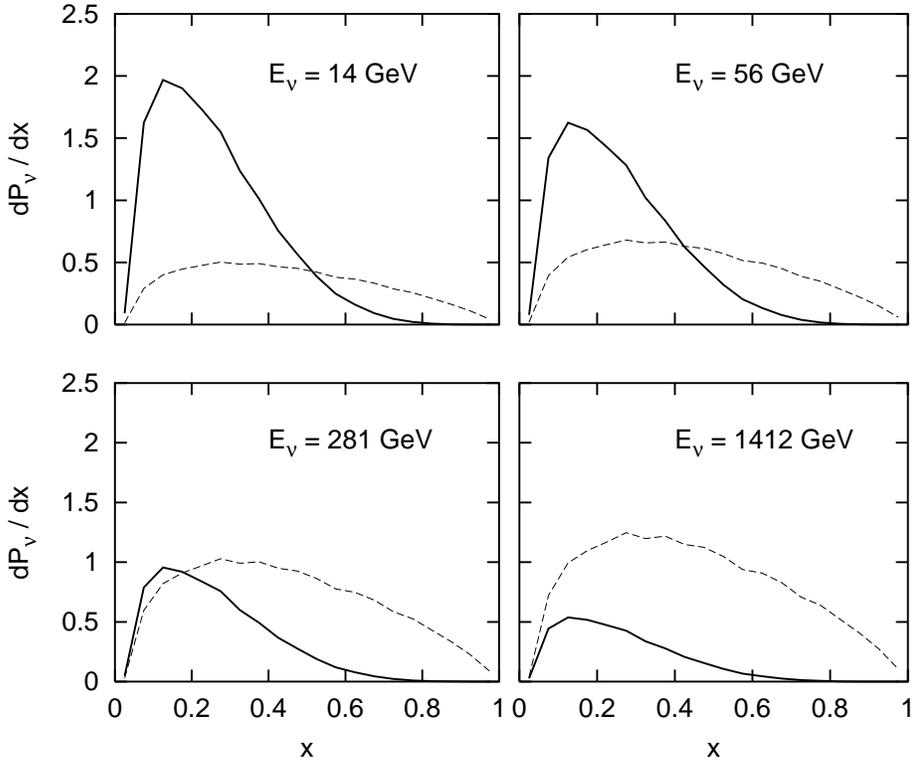,width=14cm}}
\caption{Phase space distributions for parents of $\nu_\mu+\bar{\nu}_\mu$
from decay of $\pi^\pm$ and $K^\pm$
at four values of $E_\nu$.  Solid lines are for pions and
dashed for kaons.  The curves are normalized to unity at each
energy for the sum of $\pi\rightarrow\nu$ and $K\rightarrow\nu$.
from pions and from kaons.
}
\label{PiK}
\end{figure}

\subsection{Dependence on pion and kaon production}
The dependence on the phase space for pion and kaon production
is displayed in Eq.~\ref{Zform}, which is illustrated for
several values of neutrino energy in Fig.~\ref{PiK}.
The figure illustrates the growing relative importance
of kaons.  It also illustrates the importance of the
extreme forward fragmentation region for the contribution of kaons,
reflecting the leading $p\rightarrow\Lambda + K^+$ process.  The shapes are
scale invariant in the approximations used here,
and they indicate the regions of
phase space that are important for production of atmospheric neutrinos.

A global estimate of the uncertainty due to pion and kaon production
is obtained, as in Ref.~\cite{AGLS},
by estimating the uncertainty in the Z-factors for pions and kaons
and weighting them according to Figs.~\ref{Fig1} and~\ref{resp-sk}.

\subsection{Illustration}

To obtain a simple estimate of the overall uncertainty
in the signal of upward-moving muons due to the uncertainty in the
spectrum of atmospheric neutrinos, we combine the estimates of
several sources of uncertainty as if they were independent,
statistical uncertainties.  For upward, through-going muons
in Super-Kamiokande, as shown in Fig.~\ref{resp-sk}, 
the three energy regions of the primary spectrum,
$<0.1$~TeV, $0.1<E_N < 1$~TeV and $>1$~TeV contribute
respectively fractions of 0.2, 0.4 and 0.4 to the signal.
Weighting and adding the corresponding three uncertainties 
($\pm.05,\;\pm.10,\;\pm.25$)
in quadrature gives an overall uncertainty of $\pm11$\% due to the
primary spectrum.  Uncertainties in $Z_{N\pi}$ and $Z_{NK}$ are
estimated in Ref.~\cite {AGLS} as $\pm12$\% and $\pm17$\% respectively.
Roughly equal fractions of the signal are from pions and kaons, so the
combined uncertainty from the interaction model is $\sim\pm10$\%
giving an overall estimated uncertainty of $\pm15$\%.

For upward, stopping muons in Super-K, the corresponding analysis
shows that essentially all the signal comes from primaries
with $E< 1$~TeV/nucleon, of which $\sim60$\% are in the well-measured region
between 10 and 100 GeV.  Thus the weighted uncertainty due to
uncertainty in the primary spectrum is estimated as $\pm 5$\%.
For this lower energy sample, approximately two-thirds are from decay
of pions, so the weighted uncertainty from hadronic interactions is
also somewhat smaller than for through-going muons, $\pm 10$\%, giving
a combined uncertainty of $\pm11$\%. 

Combining these two numbers results in an estimate of $\pm19$\% for
the uncertainty in the calculated ratio of stopping to through-going 
neutrino-induced,
upward muons in Super-K expected in the absence of oscillations.

\section{Summary}

The main results of the paper are contained in Eq.~\ref{Zform} and
in Eqs.~\ref{pions} and~\ref{muons} [and 
the corresponding expressions for kaons].  Eq.~\ref{Zform} displays
explicitly how the flux depends on the phase space for pion [kaon]
production.  One can see from the corresponding plots which regions
of phase space need to be measured in experiments such as HARP~\cite{HARP},
in order to improve the input to calculations of atmospheric neutrinos.

Eqs.~\ref{pions} and~\ref{muons} show respectively 
the response functions for $\nu_\mu$ and for muons.  These can be
convolved with the response of an underground detector to neutrinos
or muons to show how the signal depends on the primary spectrum.
The response functions in Fig.~\ref{Fig2} can also be used as an aid in
planning a Monte Carlo calculation of atmospheric muon or neutrino fluxes
and their signals in a particular detector.
By knowing the distribution of primary energies that contribute to
the neutrino (or muon) flux at $E_{\nu(\mu)}$, one can determine in
advance the range and distribution of primary energies needed to obtain
a result with sufficient statistical accuracy.  

Since the formulas~\ref{pions} and~\ref{muons}
display the dependence on primary energy explicitly, they can also be used
to estimate the signal from a primary spectrum of arbitrary shape
and evolution, such as might be associated with a violent solar flare.
Applications to calculation of neutrino fluxes in astrophysical beam
dumps (with appropriately modified cascade propagation factors, $G_{\pi,\mu}$)
are also possible.

\begin{figure}[htb]
\flushleft{\epsfig{figure=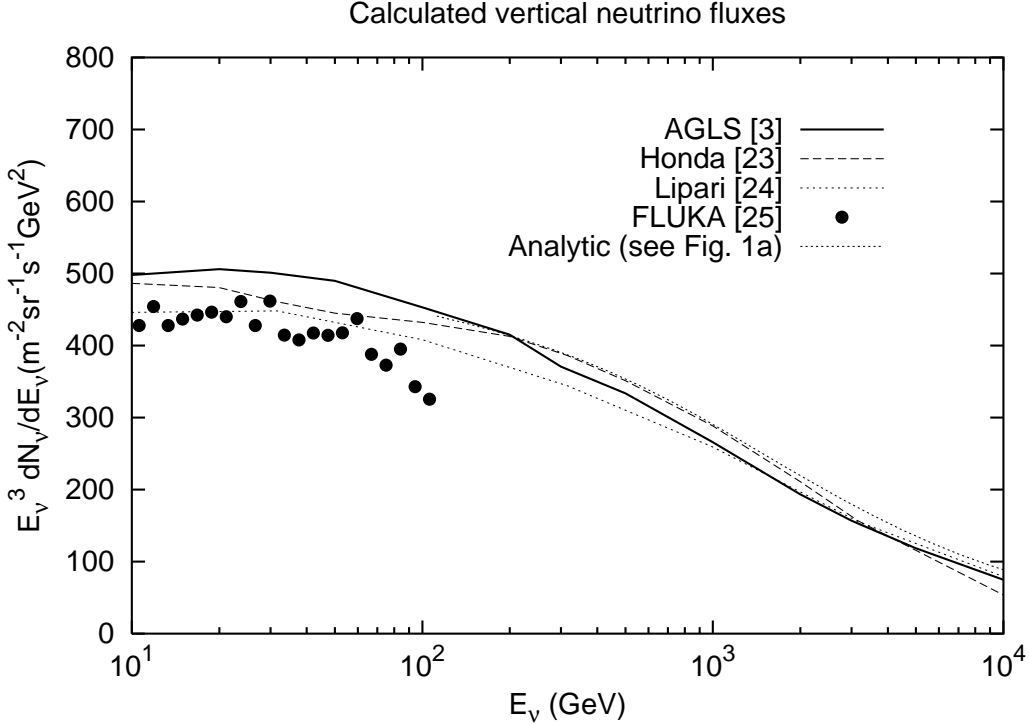,width=14cm}}
\caption{Comparison of three calculations of the vertical
flux of atmospheric neutrinos
($\nu_\mu+\bar{\nu}_\mu$)~\protect\cite{AGLS,Honda,Lipari}.
}
\label{Fig6}
\end{figure}

In \S4 we gave an example of the use of the semi-analytic approach in
estimating ``theoretical'' uncertainties in an expected neutrino-induced 
signal.  A complementary approach is simply to compare
different, independent calculations of the spectrum of atmospheric neutrinos.
Fig.~\ref{Fig6} shows a comparison
of four calculations.  The calculations of Ref.~\cite{AGLS} and~\cite{Honda}
are one-dimensional Monte Carlo calculations.  
The calculation of Ref.~\cite{Lipari}
is an analytic calculation assuming a pure power-law primary spectrum of
nucleons and scaling for the inclusive cross sections.  These three
calculations are within a range of $\approx10$\% of each other
over the relevant range of energies.  Several earlier calculations are
compared in Fig.~7 of Ref.~\cite{AGLS} and are also within this range.
The three-dimensional calculation of Battistoni {\it et al.}~\cite{Battistoni} 
is shown by the points in the figure. 
(Because of the complexity of the
3-D calculation, statistics in the near vertical bin are limited at present,
especially so at the high-energy end~\cite{FLUKA}.)  
The calculation of Ref.~\cite{Battistoni} assumed
the same primary spectrum as used in Ref.~\cite{AGLS}.  In addition,
the energy range here is well into the region for which three-dimensional
effects can be neglected.
The differences are therefore presumed to be a result of differences in the
representation of pion and kaon production.  The authors of
the calculations shown in Fig.~\ref{Fig6} are in the process of comparing
hadronic interaction models with the goal of minimizing this source
of uncertainty in the calculations of atmospheric muons and neutrinos. 

\section{Appendix: Approximate treatment of muon decay and energy loss}
For muons, which lose energy and which may decay in
the atmosphere before reaching the ground, we must first multiply
by the muon survival probability and take account of
muon energy loss between production at depth $X$
and the ground at vertical atmospheric depth $X_0\approx 1030 g/cm^2$.
This is done by multiplying Eq.~\ref{production}
by the muon survival probability before integrating over the depth, $X$, of
pion (or kaon) decay.
The muon survival probability given by Lipari~\cite{Lipari} can be
rewritten in the form
\begin{equation}
P_\mu(E_\mu,X,X_0)\;=\;
\left({X\cos\theta\over X_0}
{E_\mu\over E_\mu+\alpha(X_0/\cos\theta-X)}
\right)^{\epsilon_\mu/(E_\mu\cos\theta+\alpha\,X_0)},
\label{surv-mu}
\end{equation}
Here $E_\mu$ is the muon energy at the ground, whereas its energy at production
in Eq.~\ref{kinmu} is $E_{\mu,0}=E_\mu+\alpha(X_0/\cos\theta\,-\,X)$.

To simplify the integration, I make the approximation of replacing
$X$ in the expression for $E_{\mu,0}$ and in Eq.~\ref{surv-mu}
by $\Lambda_N\approx 120$~g/cm$^2$, which is its mean value
at low energy ($E_\mu<E_\pi\ll\epsilon_\pi\approx 115$~GeV).
Ionization energy loss is characterized by $\alpha\approx 2$~MeV/g/cm$^2$, so
$E_{\mu,0}\approx E_\mu+2$~GeV/$cos\theta$, and the approximate forms
will be valid for $E_\mu \gg 2/\cos\theta$~GeV, in practice, for
$E_\mu > 10$~GeV for the applications of this paper.

For high energy, ionization energy loss is negligible, but
radiative losses due to pair production, bremsstrahlung,
and hadronic interactions of muons begin to
become important.  However, the radiation lengths are sufficiently
large that the probability of significant
radiative energy losses in the atmosphere is small, and they
have been neglected here.

With these approximations, we can now follow the same steps used
to obtain Eq.~\ref{production2}.  The result is
\begin{eqnarray}
{dN\over dE_\mu\,dE_\pi dE_N} &=&{S_\mu(E_{\mu,0},E_\pi)\over (1-r_\pi)E_\pi}
{1\over E_\pi}F_{N\pi}(E_\pi,E_N)\\ \nonumber
&\times&{\Lambda_\pi\over\lambda_N}\,A(E_\mu)
\sum_{n=1}^\infty\,C_n(E_\mu)\,
{(-a_\pi)^{n-1}\over
1\,+\,nE_\pi\cos\theta/\epsilon_\pi}\,\phi(E_N).
\label{production3}
\end{eqnarray}
Here $$ A(E_\mu)\,=\,\left({\Lambda_\pi\cos\theta\over X_0}
{1\over 1 + 2/E_\mu\cos\theta}\right)^{\epsilon_\mu/((E_\mu\cos\theta+2)}$$
and $$
C_n\,=\,C_{n-1}\times
\left(1\,+\,{\epsilon_\mu\over (n-1)(E_\mu\cos\theta+2)}\right)
$$
with $$
C_1\,=\,\Gamma\left(1\,+\,{\epsilon_\mu\over(E_\mu\cos\theta+2)}\right)
$$

\end{document}